# Neural network concatenation for Polar Codes

Evgeny Stupachenko
*Intel Labs*
*Intel Corporation*
Santa Clara, Santa Clara
evgeny.v.stupachenko@intel.com

*Abstract* — When a neural network (NN) is used to decode a polar code, its training complexity scales exponentially as the code block size (or to be precise, as a number of message bits) increases. Therefore, existing solutions that use a neural network for polar decoders are stuck with short block sizes like 16 or 32. Despite the fact that the NN training is very complex for long polar codes, the NN decoding gives the better latency and its performance is potentially close to the maximum likelihood (ML). In this paper, we describe an efficient algorithm to create the NN decoding for a polar code of any size with the initial performance that is equal or better than that of successive cancelation (SC). Therefore, it creates an opportunity to design the NN based decoding with the performance that is as close to the ML, as the training time allows.

## I. INTRODUCTION

Polar codes are relatively new widely used error correction codes with the low complexity and the good error correction capability (improving as the code length growths). Polar codes are already a part of the 5G communication standard. Generally, a SC algorithm is used for decoding. It has very low parallelism. A NN decoder of a polar code was described in several papers. Among the first was a paper [1] where authors created a SC based partitioning and used NN decoders only for relatively short codes (with the codeword length not greater than 16). The decoding scheme became more parallel and had a better performance than the SC. However, in [1] there is no answer to the question if there is a non-brute-force method to build decoding NNs. Authors also mention that NN decoders for longer codes are possible, but very hard to train due to computational power limits (i.e. the number of codewords to be used in the training). In this paper, we propose an efficient method to create a NN decoder for polar code of any length with an initial performance not worse than the SC.

There are several ways to improve a SC algorithm's performance, like SC List [6] or improved SC [5]. In this paper, we focus on the NN based decoding only. In section II, we describe the idea of a tradeoff between the performance and the length of NN based decoders. In section III, we introduce a modification of the SC log-likelihood ratios (LLR) propagation formula, which is important for building a NN based decoder. In sections IV and V, we introduce an algorithm to build a NN to decode any polar code (define the number of layers, their activation functions and weights), section VI presents the results of the algorithm implementation.

## II. SC AND NN TRADEOFF

To understand more about the polar codes encoding, decoding and partitioning please refer to papers [1], [2] and [3]. In this paper we are using the SC algorithm recursion with each decoding task being partitioned into 2 simpler subtasks (as shown in Fig. 1) and the LLR propagation formula.

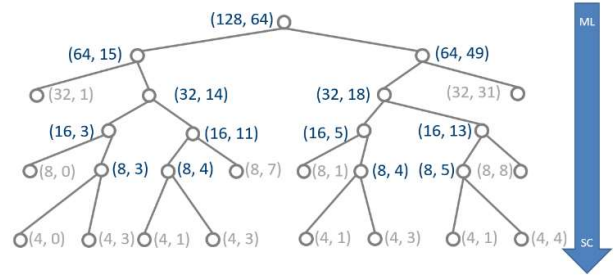

Fig. 1. SC decoding algorithm tree for (128, 64) polar code

We also rely on the binary phase shift keying modulation, an additive white Gaussian noise (AWGN) channel and the following LLR for a received symbol $y$:

$$LLR(y) = \ln\left(\frac{P(x=-1|y)}{P(x=1|y)}\right)$$

At a decoding input we have N (equal to 128 in Fig. 1 example) LLRs from a channel. To complete the decoding, we propagate the LLRs left

$$LLR_l[i] = f(LLR[i], LLR[i + \frac{N}{2}])$$

and right

$$LLR_r[i] = LLR[i] * sign(-En_l[i]) + LLR[i + \frac{N}{2}],$$

where

$$f(a,b) = -sign(a) * sign(b) * min(|a|, |b|) \quad (1)$$

and $En_l[i]$ is a codeword restored from $LLR_l[i]$. Every time LLRs are propagated left or right, we simplify the decoding, but the performance is also lost. If we propagate results to the point where subcodes are trivial: no message bits (N, 0), repetition (N, 1), parity check (N, N – 1) or only message (N, N), then we get the SC performance. However, if we stop at non-trivial codes like (16, 5), we can get the better performance by applying a NN decoder. And if we can create a NN decoder for the initial code (128, 64) we might reach the performance that is close to the ML. Unfortunately, it is very hard to create and train a NN decoder for such a long code. The NN architecture (number of layers, their dimensions and activation functions) is unknown and the number of unique tests is almost uncountable - $2^{64}$. However, for shorter subcodes like (16, 3), (16, 11), (16, 5) and (16, 13) there are known solutions to create NN decoders like in papers [1], [3] with a performance that is much better than the SC.

## III. LLR PROPAGATION FORMULA

A simplified LLR propagation formula is a key for building a NN decoder. The most widely used formula is (1). It could be replaced with typically faster (for vector processors) equivalents:

$$f(a,b) = min(max(a,b), -min(a,b))$$

$$f(a,b) = -\left|\frac{a+b}{2}\right| + \left|\frac{a-b}{2}\right|$$

$$f(a,b) = -Relu(a+b) + Relu(a-b) + b \quad (2)$$

where $Relu(x) = max(x,0)$, a regular NN activation function; "a" and "b" are LLRs. Formula (2) is a key to create a NN layer with $Relu$ activation that performs the LLR propagation.

## IV. CREATING A NN DECODER

In this section, we describe an algorithm to create a NN decoder for any polar code. The key principle is the concatenation of 2 NN based decoders into one larger NN decoder. This concatenation is based on the simplified LLR propagation formula (2) from the SC algorithm. Let's prove 3 simple lemmas:

<u>Lemma of Simplification:</u> If a NN with $L > 1$ layers contains a layer (not the last) without an activation function ($f(x) = x$), that NN could be simplified to a NN with $L - 1$ layers, without changing its functionality.

Proof: Let $A_1$ be the matrix for this layer, $b_1$ be its bias; $A_2$ be the matrix for the next layer, $b_2$ be its bias; $x$ be an input to the second layer. Then $A_1 * (A_2 * x + b_2) + b_1$ would be the output after applying these 2 layers.

$$A_1 * (A_2 * x + b_2) + b_1 = A_1 * A_2 * x + (A_1 * b_2 + b_1)$$

So, these 2 layers could be replaced by 1 layer with the matrix $A_1 * A_2$ and the bias $(A_1 * b_2 + b_1)$.

That means: when we create a NN decoder we can add layers without an activation function (to simplify an explanation) which will not change the number of layers in the final NN.

<u>Lemma of XOR equation:</u>

There are integers $a_i, b_i (i = 0..n-1)$:

$$x_0 \oplus x_1 \oplus \ldots \oplus x_{n-1} = \sum_{i=0}^{n-1} a_i * Relu\left(b_i + \sum_{j=0}^{n-1} x_j\right),$$

$x_i = \{0,1\}$, $\oplus$ is XOR (exclusive OR) operation.

Proof:

$$x_0 \oplus x_1 \oplus \ldots \oplus x_{n-1} = S_n \, mod \, 2, \text{ where } S_n = \sum_{j=0}^{n-1} x_j$$

This equation could be proved using the induction principle:

For $n = 2$ it is obvious: $x_0 \oplus x_1 = (x_0 + x_1) mod \, 2$

Now let's assume the formula is true for all values less than $n$, and prove it for $n$:

$x_0 \oplus x_1 \oplus \ldots \oplus x_{n-1} = x_0 \oplus x_1 \oplus \ldots \oplus x_{n-2} \oplus x_{n-1} = (S_{n-1} mod \, 2) \oplus x_{n-1} = ((S_{n-1} mod \, 2) + x_{n-1}) \, mod \, 2 = (S_{n-1} + x_{n-1}) mod \, 2 = S_n mod \, 2$

Now let's represent $S_n mod \, 2$ as a sum of basis functions $\sum_{i=0}^{n-1} c_i * f_i(S_n)$ where

$$f_i(x) = \begin{cases} 1, if \, x = i \\ 0, otherwise \end{cases}$$

$$c_i = i \, mod \, 2$$

Therefore, if we construct all the basis functions $f_i$ using $Relu$ – we will prove the <u>Lemma</u>. To do this let's use the induction principle again:

Basis function for $n$: $f_n(S_n) = Relu(S_n - n + 1)$

Now let's assume we have all the basis functions before $i$: $f_n, f_{n-1}, \ldots, f_{i+1}$ and create $f_i(S_n)$:

$$f_i(S_n) = Relu(S_n - i + 1) - \sum_{j=i+1}^{n} (j - i + 1) * f_j(S_n)$$

That is true because by subtracting already created $f_n, f_{n-1}, \ldots, f_{i+1}$ with corresponding multipliers from

$$Relu(S_n - i + 1) = \begin{cases} S_n - i + 1, if \, S_n > i \\ f_i(S_n), otherwise \end{cases}$$

we zero all values for $S_n > i$ and get a basis function $f_i$.

That means: XOR of any number of values could be done in just one NN layer.

In the paper we would refer to $\{-1, 1\}$ instead of $\{0, 1\}$, but the transformation between them is very simple and does not require additional layers (<u>Lemma of Simplification</u>):

$$0.5 * (x + 1): \{-1, 1\} \to \{0, 1\}$$

$$2 * x - 1: \{0, 1\} \to \{-1, 1\}$$

We need at least 1 layer to perform XOR. Otherwise there could be $a, b, c: x_0 \oplus x_1 = a * x_0 + b * x_1 + c$. This is impossible, because with different values of $x_0, x_1$ we get: $c = 0, b = 1, a = 1, a + b = 0$.

<u>Lemma of Preservation:</u> A NN layer with the $Relu$ or $Tanh$ (hyperbolic tangent) activation function can keep any input value $In$ unchanged (or with an error less than some fixed number).

Proof: For $Relu$ it will be $In + L_M$, where $L_M$ is maximum positive LLR, so that in the next layer we can subtract $L_M$ and get $In$. For $Tanh$ it will be $M * Tanh(\frac{In}{M})$, where $M$ is a big positive constant.

That means: when we construct a NN decoder, we can use input values or any output from previous layers in any layer.

Now we can proceed to a NN concatenation algorithm. Assume we have a polar code $(N, K)$ with the code word length $N = 2^n$ and $K$ message bits. Every symbol in the word is binary $\{-1, 1\}$. Any $(N > 1)$ polar code could be partitioned into 2 subcodes with the length $\frac{N}{2}$, and $K_1$, $K_2$ $(K_1 + K_2 = K)$ message bits, depending on the initial $(N, K)$ code's frozen bits selection. The tree for the partitioning will look as shown in Fig. 2.

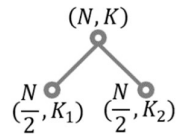

Fig. 2. Partitioning tree for the $(N, K)$ code

Let's assume we have neural networks NN$_1$ and NN$_2$ to decode the left code $(\frac{N}{2}, K_1)$ and the right code $(\frac{N}{2}, K_2)$, as in

Fig. 2. Let $LLR[0..N-1]$ be an array of LLRs received from the AWGN channel.

*A. The SC algorithm to decode the $(N, K)$ code:*

  *1) Decode the left subcode*

Compute input LLRs for the left subcode $(\frac{N}{2}, K_1)$:

$$LLR_1[i] = f(LLR[i], LLR[i + \frac{N}{2}]), \text{ where } f(a, b) \text{ is (1)}$$

and decode it using the NN$_1$ (here we get the first $K_1$ message bits).

  *2) Encode the left subcode's message bits*

Encode $K_1$ message bits to $En[0..\frac{N}{2} - 1]$

  *3) Decode the right subcode*

Swap the sign of $LLR[0..\frac{N}{2} - 1]$ elements if the encoded bit is 1:

$$LLR_D[i] = -sign(En[i]) * LLR[i],$$

compute input LLRs for the right subcode $(\frac{N}{2}, K_2)$:

$$LLR_2[i] = LLR_D[i] + LLR[i + \frac{N}{2}]$$

and decode it using NN$_2$ (here we get the last $K_2$ message bits)

Now, the idea is to create a NN to decode the code $(N, K)$ using the existing NNs: NN$_1$ and NN$_2$. An algorithm below describes how to complete this in no more than 3 additional layers. Each step from the SC algorithm description above will require an additional layer. The number of layers in the new NN will be equal to the number of layers in NN$_1$ + the number of layers in NN$_2$ + 3.

*B. Corresponding NN layers for the SC algorithm's steps*

  *1) Decode the left subcode*

As was mentioned above, the LLR propagation function (1) could be replaced with (2). That way the layer calculating inputs for the subcode $(\frac{N}{2}, K_1)$ will look as following (Fig. 3):

| E | E |
|---|---|
| E | -E |
| 0 | 0 |

Activation Relu

Fig. 3. Propagation layer 1

The last row of Fig. 3 contains bias (constants to add prior to applying the $Relu$ activation function). E is an eye matrix. With this layer we calculate

$$Relu(LLR[i] + LLR[i + \frac{N}{2}])$$

and

$$Relu(LLR[i] - LLR[i + \frac{N}{2}]).$$

Using the Lemma of Simplification for adding and the Lemma of Preservation to get $LLR[i + \frac{N}{2}]$ we compute

$$-Relu(LLR[i] + [i + \frac{N}{2}]) +$$

$$Relu(LLR[i] - LLR[i + \frac{N}{2}]) + LLR[i + \frac{N}{2}]$$

which is the LLR propagation function (2). Next, we run the NN$_1$ and get $K_1$ message bits, which we need to encode back to the $\frac{N}{2}$ code word.

  *2) Encode the left subcode's message bits*

The output from the NN$_1$ is $K_1$ message bits $\{-1, 1\}$. Using the fact that the polar encoding is just XOR of different input values and the Lemma of XOR equation, the encoding is completed in 1 layer with $Relu$ activation. Let the encoding result be $En[0..\frac{N}{2} - 1]$.

  *3) Decode the right subcode*

We need to change signs of $LLR[0..\frac{N}{2} - 1]$ according to encoded word $En[0..\frac{N}{2} - 1]$. This could be done in 1 layer too, using the same technique as in step 1 as:

$$LLR_D[i] = -sign(En[i]) * LLR[i] =$$

$$-sign(En[i]) * sign(LLR[i]) * min(|LLR[i]|, L_M) =$$

$$f(LLR[i], L_M * En[i])$$

where $f(a, b)$ is (2), $L_M$ is the maximum positive LLR and $LLR[i]$ we get using the Lemma of Preservation. Now using the Lemma of Simplification and the Lemma of Preservation we get inputs for NN$_2$:

$$LLR_2[i] = LLR_D[i] + LLR[i + \frac{N}{2}]$$

Next, we run the NN$_2$ and get $K_2$ message bits. The remaining $K_1$ message bits we get using the Lemma of Preservation. Now we get all $K=K_1 + K_2$ message bits. The problem has been solved in 3 layers.

Since decoding NNs for codes $(1, 0)$ and $(1, 1)$ are trivial, we now have an algorithm to create a decoding NN for a polar code of any size.

V. EXAMPLE

Before we jump to results, let's review a simple example for a better picture. Let's assume we need to construct a NN decoder for the code $(4, 3)$, based on a NN$_1$ decoder for the code $(2, 1)$ and a NN$_2$ decoder for the code $(2, 2)$ as in Fig. 4.

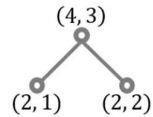

Fig. 4. Partitioning tree for the (4,3) code

| 1 | 0 | 1 | 0  | 1     | 0     | 0     | 0     |
|---|---|---|----|-------|-------|-------|-------|
| 0 | 1 | 0 | 1  | 0     | 1     | 0     | 0     |
| 1 | 0 | -1| 0  | 0     | 0     | 1     | 0     |
| 0 | 1 | 0 | -1 | 0     | 0     | 0     | 1     |
| 0 | 0 | 0 | 0  | $L_M$ | $L_M$ | $L_M$ | $L_M$ |

Activation Relu.

Fig. 5. Step 1 layer for NN concatenation

Let input LLR be $LLR[0..3]$. When we apply the layer from Fig. 5 to the $LLR[0..3]$, we get:

$$Relu(LLR[0..1] + LLR[2..3] + 0)$$

$$Relu(LLR[0..1] - LLR[2..3] + 0)$$

$$Relu(LLR[0..4] + L_M)$$

Since to count input LLRs for the $NN_1$ decoder we need just addition and subtraction:

$$LLR_1[0..1] = -Relu(LLR[0..1] + LLR[2..3]) +$$
$$Relu(LLR[0..1] - LLR[2..3]) + LLR[2..3]$$

we can insert these calculations into the $NN_1$'s first layer using the Lemma of Simplification.

| $-NN_1L_1$ | 0 | 0 |
|---|---|---|
| $NN_1L_1$ | 0 | 0 |
| $NN_1L_1$ | E | 0 |
| 0 | 0 | E |
| $NN_1C$ | C | C |

$NN_1$ activation

Fig. 6. $NN_1$'s first layer

In Fig. 6 $NN_1L_1$ – is the first layer of $NN_1$, $NN_1C$ – are $NN_1$'s first layer biases, C – are biases required to preserve values (see the Lemma of Preservation). Next, we apply the rest $NN_1$ layers and get $K_1 = 1$ message bit $I[0]$, preserving $LLR[0..3]$. As the message bit is only 1, we don't need to encode it. However, in general case it will take us 1 layer with Relu activation to complete any binary encoding (see the Lemma of XOR equation). Now to get input LLRs for the $NN_2$ decoder we need to multiply $LLR[0..1]$ by sign of $I[0]$ and subtract the result from $LLR[2..3]$.

| $L_M$ | $L_M$ | $-L_M$ | $-L_M$ | 0 | 0 | 1 |
|---|---|---|---|---|---|---|
| 1 | 0 | 1 | 0 | 0 | 0 | 0 |
| 0 | 1 | 0 | 1 | 0 | 0 | 0 |
| 0 | 0 | 0 | 0 | 1 | 0 | 0 |
| 0 | 0 | 0 | 0 | 0 | 1 | 0 |
| 0 | 0 | 0 | 0 | $L_M$ | $L_M$ | $L_M$ |

Activation Relu

Fig. 7. Step 3 layer for NN concatenation

With the input to the layer in Fig. 7: $I[0]$ and $LLR[0..3]$ we get:

$$Relu(LLR[0..1] + L_M * I[0..1])$$
$$Relu(LLR[0..1] - L_M * I[0..1])$$
$$\text{preserved } I[0] \text{ and } LLR[2..3]$$

Now using the Lemma of Simplification we get input LLRs for the $NN_2$ decoder:

$$LLR_2[0..1] = LLR[2..3] - Relu(LLR[0..1] + L_M * I[0]) +$$
$$Relu(LLR[0..1] - L_M * I[0]) + L_M * I[0]$$
$$= LLR[2..3] - LLR[0..1] * sign(I[0])$$

Using the same technique as shown in Fig. 6, we apply the $NN_2$ decoder and get the remaining $K_2 = 2$ message bits.

## VI. RESULTS

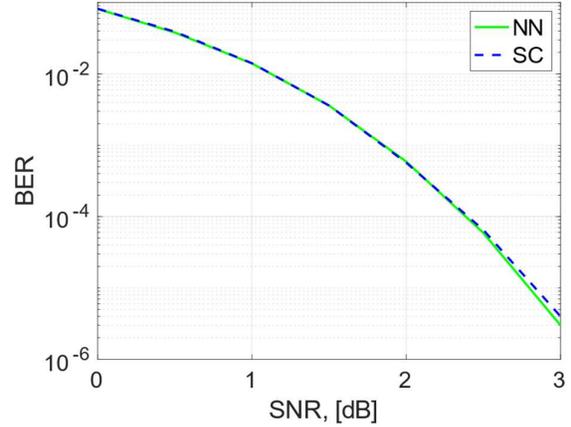

Fig. 8. Initial performance of constructed (8, 4) + (8, 7) NN

Fig. 8 shows different signal to noise ratio's (SNR) bit error rate (BER) for the NN based decoding and the SC decoding, as applied to polar code (16, 11). The NN based decoding is built from the concatenation of the (8, 4) and the (8, 7) NN decoders. The curves confirm that the initial performance of the NN concatenation is not worse than the SC.

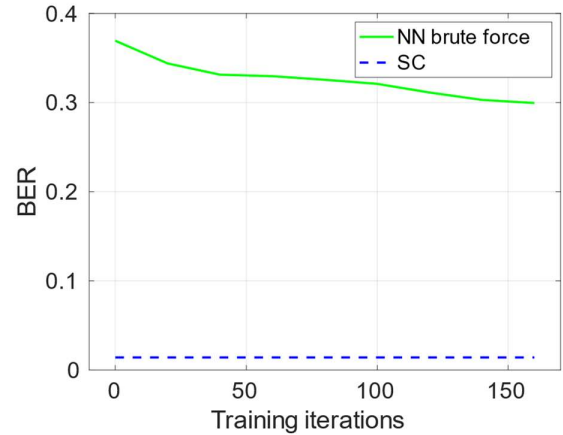

Fig. 9. NN concatenation performance with unknown weights

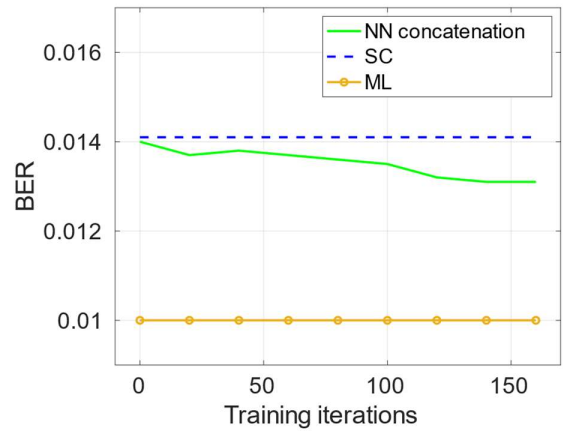

Fig. 10. NN concatenation performance with prebuilt weights

Fig. 9 shows the BER for (16, 11) NN based decoder, when we do not know NN weights and start a training from scratch. The chart covers 160 iterations of training. SC – is the successive cancelation's BER. The SNR is fixed and equal to 1. As we can see we are not even close to the SC's BER and the training speed is low.

Fig. 10 shows how (16, 11) NN based decoder built as the concatenation of NN decoders (8, 4) and (8, 7) consistently improve the BER starting from the SC's BER and moving to the ML's BER. The NN architecture (number of layers/their dimensions and activation functions) is the same for both Fig. 9 and Fig 10.

The creative part here is to reduce the number of tests to speed up the training process. To achieve this, we manually create a set of tests that are wrongly decoded using the NN concatenation, but successfully decoded using the ML metric. Generally, to train a polar decoding NN we need about $10^6$ tests for each codeword. For code (16, 11) it will be $10^6 * 2^{11}$, taking too long to train. Moreover, since for larger codes we need NNs with more layers, each training step takes longer too. For code (16, 11) we were unable to achieve the SC performance when train it from scratch.

As a benefit, the LLR propagation formula (1) replacement speeds up current implementations of the SC (especially for vector processors) by 5-20% as measured on Intel and Tensilica processors.

## VII. Conclusion

In this paper, we proposed an algorithm to create a NN polar code decoder for any code length, which is a step forward, as compared to papers [1] and [3]. Moreover, this creates an opportunity to further reduce the decoding latency and allow switching from one polar code to another with just replacing NN weights.

The future work is to choose the most efficient training tests subset for the NN concatenation, to reduce the number of layers in the NN concatenation and to prune existing concatenation layers.